\documentclass[twocolumn,showpacs,preprintnumbers,amsmath,amssymb]{revtex4}

\def\be{\begin{equation}}
\def\ee{\end{equation}}
\def\bee{\begin{eqnarray}}
\def\eee{\end{eqnarray}}

\input{epsf}

\begin{document}
\title{The toroidal momentum pinch velocity}

\author{A.G. Peeters, C. Angioni, D. Strintzi}

\address{Max Planck Institut fuer Plasmaphysik, EURATOM association, 
Boltzmannstrasse 2 85748 Garching, 
Germany}

\begin{abstract}
In this letter a pinch velocity of toroidal momentum is shown to exist for the 
first time. 
Using the gyro-kinetic equations in the frame moving with the equilibrium 
toroidal velocity, it is shown that the physics effect can be elegantly formulated 
through the ``Coriolis'' drift. A fluid model is used to highlight the main
coupling mechanisms between the density and temperature perturbations on the 
one hand and the perturbed parallel flow on the other. Gyro-kinetic calculations 
are used to accurately asses the magnitude of the pinch. The pinch velocity leads
to a radial gradient of the toroidal velocity profile even in the absence of a 
torque on the plasma. It is shown to be 
sizeable in the plasmas of the International Thermonuclear Experimental Reactor 
(ITER) leading to a moderately peaked rotation profile. 
Finally, the pinch also affects the interpretation of current experiments. 
\end{abstract}

\pacs{52.25.Fi, 52.25.Xz, 52.30.Gz, 52.35.Qz, 52.55.Fa}

\maketitle


In a tokamak the total toroidal angular momentum is a conserved quantity
in the absence of an external source. 
Transport phenomena determine the rotation profile which is of interest 
because a radial gradient in the toroidal rotation is connected with an 
ExB shearing that can stabilise micro-instabilities
\cite{BIG90,WAL94,HAH95} and, hence, improve confinement. 
Furthermore, a toroidal rotation of sufficient magnitude can stabilise
the resistive wall mode \cite{BON94,STR95,REI06}. 
In present day experiments the rotation is often determined
by the toroidal torque on the plasma that results from the neutral beam 
heating. 
Such a torque will be largely absent in a reactor and it is 
generally assumed that the rotation, and hence its positive influence, 
will be small. 
The novel pinch velocity described in this letter, however, may generate
a sizeable toroidal velocity gradient in the confinement region even in 
the absence of a torque.

We will focus on the Ion Temperature Gradient (ITG) mode, which
is expected to be the dominant instability governing the ion heat channel 
in a reactor plasma. 
The equations are formulated using the gyro-kinetic framework 
\cite{FRI82,DUB83,LEE87,HAH88}, which has been proven successful in explaining 
many observed transport phenomena 
\cite{BOU02,SYN02,CAN03,GAR03,ERN04,ROM04,KIN05,PEE05a,JEN05,ANG05,PEE05b,BOT06}. 
Because of the rotation, the background electric field cannot be ordered
small \cite{HAH92,ART94,HAH96,BRI95}, and the starting point for the derivation is 
a set of equations for the time evolution of the guiding centre ${\bf X}$ 
and the parallel (to the magnetic field) velocity component ($v_\parallel$) in the 
co-moving system (with background velocity ${\bf u}_0$) obtained from Ref.~\cite{BRI95} 
\be
{{\rm d}{\bf X} \over {\rm d} t} = v_\parallel {\bf b} + 
{{\bf b} \over e B_\parallel^*}  \times  ( e \nabla \phi + \mu \nabla B 
+ m {\bf u}_0^* \cdot \nabla {\bf u}_0^* ),
\ee
\be
{{\rm d} v_\parallel \over {\rm d} t} = - {{\bf B}^* \over m B_\parallel^*} \cdot
(e \nabla \phi + \mu \nabla B + m {\bf u}_0^* \cdot \nabla {\bf u}_0^* ).
\ee
Here ${\bf b} = {\bf B}/B$ is the unit vector
in the direction of the magnetic field (${\bf B}$), $\phi$ is the perturbed 
gyro-averaged potential (i.e. the part not connected with the 
background rotation), $\mu$ the magnetic moment, $m$ ($e$) the particle mass 
(charge), and ${\bf u}_0^* = {\bf u}_0 + v_\parallel {\bf b}$. 
For the background velocity (${\bf u}_0$) we assume a constant rigid body toroidal rotation 
with angular frequency ${\bf \Omega}$ (this is an equilibrium solution see, for instance, 
Refs. \cite{HIR81,BRI95,PEE98})
\be 
{\bf u}_0 = {\bf \Omega} \times {\bf X} = R^2 \Omega \nabla \varphi,
\ee
where $\varphi$ is the toroidal angle.
We briefly outline the derivation of the final equations here. 
More details can be found in \cite{PEE07}. 
The background velocity ${\bf u}_0$ will be assumed smaller than the thermal 
velocity, and only the terms 
linear in ${\bf u}_0$ will be retained. This eliminates the centrifugal forces but 
retains the Coriolis force.
Furthermore the low beta approximation is used for the equilibrium magnetic field 
(i.e. ${\bf b} \cdot \nabla {\bf b} \approx \nabla_\perp B / B$ where $\perp$ indicates the 
component perpendicular to the magnetic field). With these assumptions
\be 
{\bf u}_0^* \cdot \nabla {\bf u}_0^* \approx v_\parallel^2 {\nabla_\perp B \over B} 
+ 2 v_\parallel {\bf \Omega} \times {\bf b}.
\ee
Using the definition of ${\bf B}^*$ (see Ref. \cite{BRI95}) and expanding up to 
first order in the normalised Larmor radius $\rho^* = \rho / R$, where $R$ is 
the major radius, one obtains 
\be 
{\bf B}^* ={\bf B} + {B \over \omega_c} \nabla \times {\bf u}_0^* = 
B \biggl [ {\bf b} + {2 {\bf \Omega} \over \omega_c} + 
{v_\parallel \over \omega_c} {{\bf B} \times \nabla B \over B^2} \biggr ] 
\ee 
and $B_\parallel^* = {\bf b} \cdot {\bf B}^* = B(1 + 2 \Omega_\parallel / \omega_c)$ 
($\omega_c = e B / m$ is the gyro-frequency). Expanding
now the equations of motion retaining only terms up to first order in $\rho^*$ 
yields 
\be 
{{\rm d} {\bf X} \over {\rm d} t} = v_\parallel {\bf b} + {{\bf b} \times 
\nabla \phi \over B} + {v_\parallel^2 + v_\perp^2 / 2 \over \omega_c } 
{{\bf B } \times \nabla B \over B^2}+ 2 {v_\parallel \over \omega_c} {\bf 
\Omega}_\perp
\label{dxdtfinal}
\ee
The terms in this equation are from left to right, the parallel motion 
($v_\parallel {\bf b}$), the ExB velocity ${\bf v}_E$, the combination of 
curvature and grad-B drift ${\bf v}_d$, and 
an additional term proportional to ${\bf \Omega}_\perp$. 
An interpretation of this term can be found if one uses the standard expression
for a drift velocity (${\bf v}_D$) due to a force (${\bf F}$) perpendicular to the 
magnetic field ${\bf v}_D = {\bf F} \times {\bf B} / eB^2$. 
Substituting the Coriolis force ${\bf F}_c = 2 m {\bf v} \times \Omega$, and taking 
for the velocity (${\bf v}$) the lowest order (parallel) velocity one obtains
\be 
{\bf v}_{dc}= {{\bf F}_c \times {\bf B} \over e B^2} = {2v_\parallel \over 
\omega_c } {\bf \Omega}_\perp
\ee
The last term in Eq.~(\ref{dxdtfinal}) is therefore the Coriolis drift. 
Expanding the terms in the equation for the parallel velocity to first order 
in $\rho^*$ one can derive 
\be 
m v_\parallel {{\rm d} v_\parallel \over {\rm d} t} = -e {{\rm d} {\bf X} 
\over {\rm d} t} \cdot \nabla \phi - \mu {{\rm d} {\bf X} \over {\rm d} t} 
\cdot \nabla B
\ee
where ${\rm d} {\bf X} / {\rm d} t$ is given by Eq.~(\ref{dxdtfinal}). The 
derived equations are similar to the non-rotating system, with the difference 
being the additional Coriolis drift. It follows that this Coriolis drift 
appears in a completely symmetric way compared with the curvature and grad-B drift. 
  
In this letter the approximation that assumes circular surfaces and small 
inverse aspect ratio ($\epsilon$) is used. In this case the Coriolis drift 
adds to the curvature and grad-B drift 
\be 
{\bf v}_d + {\bf v}_{dc} \approx {v_\parallel^2 + 2 v_\parallel R \Omega  + 
v_\perp^2/2 \over \omega_c R } {\bf e}_z,
\ee
where ${\bf e}_z$ is in the direction of the symmetry axis of the tokamak. 
The linear gyro-kinetic equation is solved using the ballooning transform 
\cite{CON78}. The equations, except from the Coriolis drift are standard and 
can be found in, for instance, Ref.~\cite{KOT95}. 
In the following $u^\prime \equiv - R \nabla R\Omega / v_{th}$ and 
$u \equiv R\Omega / v_{th}$. Unless explicitly stated otherwise
all quantities will be made dimensionless 
using the major radius $R$, the thermal velocity $v_{th} \equiv \sqrt{2 T / m_i}$,
and the ion mass $m_i$. Densities will be normalised with the electron density. 
The toroidal momentum flux is approximated by the flux of 
parallel momentum ($\Gamma_\phi$) which is sometimes normalised with the 
total ion heat flow ($Q_i$)
\be 
(\Gamma_\phi,Q_i) = \biggl \langle {\bf v}_E \int {\rm d}^3 {\bf v} 
\biggl (m v_\parallel, {1\over 2} m v^2 \biggr ) f \biggr \rangle ,
\ee
where $f$ is the (fluctuating) distribution function and the brackets denote the 
flux surface average.

Before turning to the gyro-kinetic calculations, first the implications of the 
Coriolis drift will be investigated using a simple fluid model
(more extended models have been published in Refs.~\cite{WEI89,WAL97}). 
A (low field side) slab like geometry will be assumed with all plasma parameters 
being a function of the x-coordinate only. The magnetic field is  
${\bf B} = B{\bf e}_y$, $\nabla B = -{B/R}{\bf e}_x$,
The model can be build by taking moments of the gyro-kinetic equation in 
(${\bf X},v_\parallel, v_\perp$) coordinates
\be 
{\partial f \over \partial t} + ({\bf v}_d + {\bf v}_{dc})\cdot \nabla f = - 
{\bf v}_E \cdot \nabla F_M  - {e F_M\over T }  ({\bf v}_d + {\bf v}_{dc} ) 
\cdot \nabla \phi,
\label{dfdtequation}
\ee
where $F_M$ is the Maxwell distribution. Note that translation symmetry in the 
z-direction is assumed, eliminating the parallel dynamics. 
Building moments of these equations neglecting the heat fluxes 
(this a clear simplification, see for instance \cite{DOR93,HAM93,BEE96,SCO05}), 
and taking the space and time dependence of the perturbed quantities as 
$\exp[ik_\theta z - i \omega t]$, one arrives at the following equations for 
the perturbed density ($n$) normalised to the background density ($n_0$), the 
perturbed parallel velocity ($w$) normalised with the thermal velocity, and 
the perturbed ion temperature ($T$) normalised with the background ion 
temperature ($T_0$)
\be 
\omega n + 2 (n + T) + 4 u w = \biggl [ {R\over L_N} - 2 \biggr ] \phi,
\ee
\be 
\omega w + 4w + 2 u n + 2 u T   = [ u^\prime - 2 u] \phi ,
\ee
\be 
\omega T + {4 \over 3} n + {14\over 3} T +{8\over 3}uw = 
\biggl [ {R\over L_T } -{4\over 3} \biggr ]\phi.
\ee
Here $R/L_N \equiv - R \nabla n_0 / n_0$, $R/L_T \equiv - R \nabla T_0 / T_0$, 
the potential $\phi$ is normalised to $T_0/e$, and the frequency is normalised 
with the drift frequency $\omega_D = - k_\theta T_0 / e B R$. 
The Coriolis drift (all terms proportional to $u$) 
introduces the perturbed velocity in the equations for the perturbed density, 
and temperature. 
However, since  $u \ll 1$ the influence of the Coriolis drift on the ``pure'' 
ITG (with $u=0$) is relatively small.
The Coriolis drift generates a coupling between $w$ and the density, 
temperature as well as potential fluctuations. 
Note that for $u=0$ the perturbed velocity is directly related
to the gradient $u^\prime$, resulting in a purely diffusive flux.  
For finite rotation ($u\ne 0$) the ITG will generate a perturbed parallel 
velocity $w$, which is then transported by the perturbed ExB velocity. If the 
perturbed temperature is kept the expressions for the 
momentum flux become rather lengthy and are, therefore, reported elsewhere
\cite{PEE07}. Retaining only the coupling with the perturbed density and 
potential, and assuming an adiabatic electron response 
($n = \phi / \tau$ with $\tau = T_e/ T_0$ being the electron to ion 
temperature ratio) one can derive
\be 
\Gamma_\phi =  {1\over 4} k_\theta \rho {\rm Im} [ \phi^\dagger w] = \chi_\phi \biggl [ 
u^\prime - {2+2\tau \over \tau} u \biggr ] \label{fluidpinch}, 
\ee
with 
\be 
\chi_\phi = - {1\over 4} k_\theta \rho {\gamma \over (\omega_R + 4)^2 + \gamma^2} \vert \phi \vert^2.
\ee
Here, the dagger denotes the complex conjugate, $\omega_R$ is the real
part of the frequency, and $\gamma$ the growth rate of the mode. 
Note that $\chi_\phi$ is positive since $\omega_R$ ($\gamma$) are normalised to 
$\omega_D = - k T_0 / e B R$.
The second term between the square brackets of Eq.~(\ref{fluidpinch}) represents an inward pinch
of the toroidal velocity (the word pinch is used here because the flux is proportional to $u$, unlike
off-diagonal contributions that are due to pressure and temperature gradients \cite{COP02,PEE06})
If one assume no torque, i.e. $\Gamma_\phi = 0$ it can be seen that the pinch can lead to a sizeable 
gradient length $R/L_u \equiv R \nabla u / u  = 4$ (for $\tau = 1$). The peaking is in roughly the 
same range as the expected density peaking \cite{ANG03}.  
\begin{figure}[htbp]%
  \begin{center}%
     \setlength{\unitlength}{1.0cm}%
     \begin{picture}(8,5)%
        \epsfxsize=7.5cm%
        \epsffile{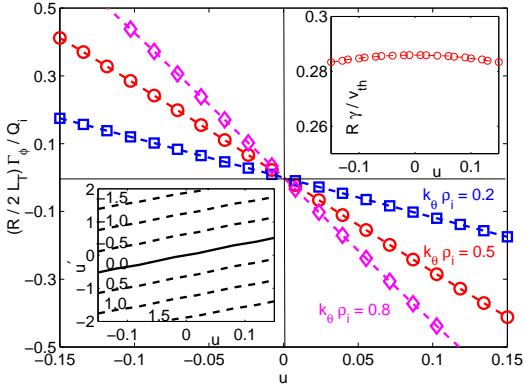}%
     \end{picture}%
  \end{center}%
  \caption{$(R/2L_T) \Gamma_\phi / Q_i$ as a function of $u$ for three values of $k_\theta 
\rho_i$ 0.5 (o), 0.2 (squares), and 0.8 (diamonds). The top right graph shows 
the growth rate as a function of $u$ and the down left graph the contour lines of 
$(R/2L_T) \Gamma_\phi / Q_i$ as a function of $u$ and $u^\prime$, both for $k_\theta \rho_i = 0.5$.
In the latter graph the thick line denotes zero momentum flux, i.e. the stationary point for 
zero torque}%
  \label{vcorscan}%
\end{figure}%

Fig.~\ref{vcorscan} shows the parallel momentum flux as a function of the toroidal velocity $u$ 
obtained from linear gyro-kinetic calculations using the LINART code \cite{PEE04} (in which unlike 
Eq.~(\ref{dfdtequation}) the parallel dynamics is kept) for three different values of the poloidal 
wave vector ($k_\theta \rho_i = 0.2$, 0.5, and 0.8).
The parameters of each of the gyro-kinetic calculations in this 
letter are those of the Waltz standard case \cite{WAL95}: $q=2$, magnetic shear $\hat s = 1$, 
$\epsilon = 0.1$, $R/L_N = 3$, $R/L_T = 9$, $\tau =1$, $u=u^\prime = 0$. In the presented
scans one of these parameters is varied while keeping the others fixed. 
Since the flux from Fig.~\ref{vcorscan} is linear in the velocity, 
a constant pinch velocity exists in agreement with the fluid model. 
The influence of the toroidal velocity on the growth rate is small. 
The bottom left graph shows the contour lines of the flux as 
a function of $u$ and $u^\prime$. The fact that the contour lines are straight
means that the momentum flux is a linear combination of the diffusive part 
($\propto \chi_\phi u^\prime$) and the pinch velocity ($V_\phi u$) 
\be 
\Gamma_\phi = [\chi_\phi u^\prime + V_\phi u ]
\ee
The diagonal part has been calculated previously using fluid \cite{MAT88,ITO92,DOM93,DIA94,GAR02} 
as well as gyro-kinetic theory \cite{PEE05,KIN05b}.
The pinch velocity is negative (inward) for positive $u$ such that it enhances the gradient.
It changes sign with $u$ such that for negative velocities it will make $u^\prime$ more 
negative, i.e. the pinch always enhances the absolute value of the velocity gradient in 
agreement with the results from the fluid theory. 
Fig.~\ref{vcorscan} also shows that the pinch decreases with $k_\theta \rho_i$. It is noted 
here that also ${\chi_\phi}$ in becomes smaller for smaller $k_\theta \rho_i$
\cite{PEE06}.

\begin{figure}[htbp]%
  \begin{center}%
     \setlength{\unitlength}{1.0cm}%
     \begin{picture}(9,5)%
        \epsfxsize=6.5cm%
        \epsffile{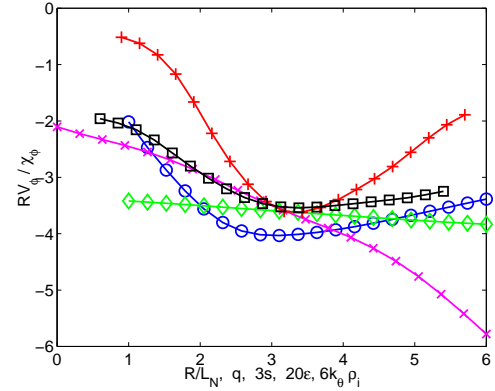}%
     \end{picture}%
  \end{center}%
  \caption{$R V_\phi / \chi_\phi$ as a function of various parameters: $R/L_N$ (x), $3 \hat s$ (+),
$q$ (o), and  $20 \epsilon$ (diamonds), and $6k_\theta \rho_i$ (squares)}%
  \label{allscans}%
\end{figure}%

Fig.~\ref{allscans} shows the normalised pinch velocity $R V_\phi / \chi_\phi$ as a function of 
various parameters. 
The magnetic shear and the density gradient have a rather large impact. 
Note that both due to $\hat s$, as well as due to $q$, $R/L_N$ and $\epsilon$, the pinch 
velocity is expected to be small in the inner core, but sizeable in the confinement region. 

The novel pinch velocity described in this letter has several important consequences. 
It can explain a gradient of the toroidal velocity in the confinement region of the plasma 
without momentum input. 
A spin up of the plasma column without torque has indeed been observed 
\cite{ERI97,RIC99,HUT00,RIC04,GRA04,SCA06}. 
Although a consistent description ordering the different observations is still lacking,
the calculations of this letter show that the pinch velocity is expected to play an important role. 
This finite gradient without torque is especially important for a tokamak reactor in which the 
torque will be relatively small. 
From the calculations shown above, and for typical parameters in the confinement region of a 
reactor plasma, one obtains a gradient length $R/L_u=u^\prime / u$ in the range 2-4 representing 
a moderate peaking of the toroidal velocity profile similar to that of the density. 
Unfortunately, the current calculation only yields the normalised toroidal velocity gradient. 
In order to determine the velocity gradient one would need to know the edge rotation velocity. 
This situation is similar to that of the ion temperature \cite{KOT95b}. 

The existence of a pinch can resolve the discrepancy between the calculated 
$\chi_\phi$ and the experimentally obtained effective diffusivity 
($\chi_{\rm eff} = \Gamma_\phi / u^\prime$). 
The latter is often found to decrease with increasing minor radius and to be smaller than 
the theoretical value of $\chi_\phi$ in the outer region of the plasma \cite{NIS05,VRI06,ANG07}. 
The pinch indeed leeds to a decrease of $\chi_{\rm eff}$
\be 
\chi_{\rm eff} = \chi_\phi \biggl [ 1 + {R V_\phi \over \chi_\phi} {1 \over R/L_u} \biggr]. 
\ee
The calculations in this letter show that the second term in the brackets can be of the order -1, 
leading to $\chi_{\rm eff} < \chi_i$. 



\bibliographystyle{aip}



\end{document}